\documentstyle[12pt,colordvi,epsfig]{article}
\newcommand{\be}{\begin{eqnarray}}
\newcommand{\ee}{\end{eqnarray}}
\newcommand{\ud}{\underline}
\begin{document}
\begin{center}
\medskip
\medskip
\medskip          
\Large{\bf $J/\psi$ Gluonic Dissociation Revisited : I.\\
Fugacity, Flux And Formation Time Effects}
\vskip 0.2in
\large{ B. K. Patra$^1$ and V. J. Menon$^2$}
\vskip 0.2in
\normalsize{$^1$ Dept. of Physics, Indian Institute of Technology,
Roorkee 247 667, India\\
$^2$ Dept. of Physics, Banaras Hindu University, 
Varanasi 221 005, India}
\end{center}
\vskip 0.2in

\begin{center}
Abstract
\end{center}

We revisit the standard treatment [Xu, Kharzeev, Satz and Wang, Phys. Rev.
C {\bf 53}, 3051 (1996)] of $J/\psi$ suppression due to gluonic bombardment
in an equilibrating quark-gluon plasma. Effects arising from gluon fugacity,
relative $g-\psi$ flux, and $\psi$ meson formation time are correctly
incorporated in the formulation of the gluon number density, 
velocity-weighted cross section, and the survival probability. Our new
formulae are applied to numerically study the pattern of $J/\psi$
suppression in the central rapidity region at RHIC/LHC energies. The 
temperature and transverse momentum dependence of our graphs
have noticeable  differences from those of Xu et al.

\vskip 0.3in

\newpage
\section{INTRODUCTION} 
Relativistic heavy ion collision experiments at CERN SPS/LHC 
and BNL/RHIC are believed to have led to a phase transition
from the hadronic world into deconfined and/or chirally symmetric
state of free quarks and gluons, the so called quark-gluon plasma
(QGP)~\cite{review}-\cite{GH}. However, as of now, no conclusive evidence of QGP 
formation has 
been discerned. Among the most hotly debated, theoretically proposed
signatures in this context are the erstwhile $J/\psi$ suppression due to
medium influence~\cite{GH} and recent $J/\psi$ enhancement via dynamical
regeneration~\cite{Thews}. The well known mechanisms responsible for $J/\psi$
suppression are summarized briefly in the Appendix for the sake
of completeness.

Attention in the sequel will be focused on the break-up of the
$J/\psi$ owing to bombardment with energetic gluons~\cite{khar}. For this
mechanism, Xu et al.~\cite{xu} employed statistical mechanics coupled with
phenomenological QCD to calculate the $J/\psi$ survival probability in a
temporally evolving parton gas. The aim of the present paper is to
extend/modify the work of Xu et al in the following multifold respects :

(i) {\underline{\bf Gluon fugacity effect}}: For large momentum gluons which
are responsible for the $\psi$ meson dissociation Xu et al.~\cite[Eq.(7)]{xu} 
replaced the gluon fugacity $\lambda_g$ by unity in the denominator of the 
Bose-Einstein distribution function. However, in the
early stage of evolution the system may be quite far from chemical
equilibrium implying that $\lambda_g$ need not be close to unity.
In Sec. 2 below we derive a new formula for the gluon number density
$n_g$ valid for general $\lambda_g \le 1$.

(ii) {\underline{\bf Relative flux effect}}: Xu et al.~\cite[Eq.(8)]{xu}
were interested in the product $\Gamma = v_{\rm{rel}} \sigma$ where the
gluon-$\psi$ break-up cross section $\sigma$ was written in the $\psi$
meson rest frame, but unfortunately their relative flux $v_{\rm{rel}}$
was evaluated in the fireball frame. In Sec. 3 below we modify
this procedure by treating the product $\Gamma$ strictly in the
$\psi$ rest frame.

(iii){\underline{\bf Formation time effect}}: For computing their survival 
probability $S(p_T)$ Xu et al.~\cite[Eq.(14)]{xu} used an integration
over $\tau_\psi$ having lower limit $0$, where $\tau_\psi$ is the 
proper time measured in $J/\psi$ rest frame. This is very inconvenient
because the gluon density $n_g(t)$ and the thermal-averaged 
cross section $\langle v_{\rm{rel}} \sigma \rangle$ are natural
functions of the usual time $t$ in fireball rest frame. In Sec. 4
below we write a modified expression for $S(p_T)$ using $t$ integration
where formation times of the QGP as well as $J/\psi$ are
explicitly included. Of course, in our numerical results in Sections
2, 3, 4 the explicit velocity profiles of hydrodynamic flow are ignored.
Finally, our main conclusions appear in Sec. 5.

\section{Number Density}
\subsection{\underline{Preliminaries}}

Assuming thermal equilibrium and working in the fireball rest frame
let the symbol $T$ denote the absolute temperature, $K$ the
gluon four momentum, $16$ the spin-colour degeneracy factor,
$\lambda_g \le 1$ the gluon fugacity, and $f=f(K^0, T, \lambda_g)$
the one-body gluon distribution function. Then the gluon number density $n_g$ 
is obtained from 
\be
n_g = 16 \int\frac{d^3K}{{(2\pi)}^3} f = \frac{8}{\pi^2}
\int_0^\infty dK^0 {K^0}^2 f
\ee

\subsection{\underline{Xu Procedure}}
For near chemical equilibration Xu et al.~\cite[Eq.(7)]{xu} employed an
approximate, factorized Bose-Einstein distribution
\be
f^{\rm{Xu}} = \frac{\lambda_g}{e^{K^0/T}-1} = \lambda_g \sum_{n=1}^\infty
e^{-nK^0/T}
\ee
which led them to a number density depending on the fugacity
linearly through
\be
n_g^{\rm{Xu}} = \frac{16}{\pi^2} T^3 \lambda_g \zeta(3)
\ee

\subsection{\ud{Our Proposal}} In order to tackle the possibility of gluon
chemical non-equilibration we use the full Bose-Einstein form
\be
f^{\rm{Our}} = \frac{\lambda_g}{e^{K^0/T}-\lambda_g} = \sum_{n=1}^\infty
\lambda_g^n e^{-nK^0/T}
\ee
which guides us to a number density containing the fugacity in a power
series via
\be
n_g^{\rm{Our}} = \frac{16}{\pi^2} T^3 \sum_{n=1}^\infty 
\frac{\lambda_g^n}{n^3}
\ee
Remembering that $1/n^3$ type series converges rapidly with $n$, numerical
comparison of Eqs. (3, 5) is easily done via the ratio
\be
\frac{n_g^{\rm{Xu}}}{n_g^{\rm{Our}}} \sim (1 +\frac{1}{8})/
(1+\frac{\lambda_g^2}{8})
\ee
which, however, requires the knowledge of $\lambda_g$ at various
times. Of course, the algebraic reason for the inequality
$n_g^{\rm{our}} < n_g^{\rm{Xu}}$ is the fact that the 
distribution function $f^{\rm{Our}} < f^{\rm{Xu}}$ as long
as $\lambda_g < 1$. In other words, the effect of correct
fugacity (before chemical equilibration) is to {\em reduce} the number
density of gluons below the value of Xu et al.

\subsection{\ud{Initial Conditions}} It is well recognized that the
scenario resulting from relativistic heavy-ion collisions is rapidly 
time-dependent. Quick scattering among the partons drives the QGP
to thermal equilibrium  in the fireball rest frame within a time
$t_i = \tau_0 \sim 1/\Lambda \sim 0.5 \rm{fm}/c$ 
where the suffix $i$ stands for ``initial" and $\Lambda$ is the
QCD energy scale. The initial conditions predicted by HIJING
Monte Carlo simulation are summarized in Table 1. There gluon densities
computed via Xu procedure (Eq.3) and our proposal (Eq. 5) are also
listed. Clearly the {\em{relative}} difference between $n_{gi}^{\rm{Xu}}$
and $n_{gi}^{\rm{Our}}$ is of the order of $1/8 \sim 12 \%$
which is significant.
\begin{table}[h]
\caption{Initial values for the temperature, time, fugacities etc. at RHIC(1), LHC(1) only~\protect\cite{hijing}}
\begin{tabular}{lcccccc}
\hline
& $T$ (GeV) & $t_i=\tau_0$ (fm) & $\lambda_g$ & $\lambda_q$ & 
$n_g^{\rm{Xu}}{\rm{(fm)}}^{-3}$ & $n_g^{\rm{Our}}{\rm{(fm)}}^{-3}$ \\
   &  & & & & &\\
RHIC(1)  & 0.55 & 0.70 & 0.05 & 0.008 & 2.11 & 1.76\\
  &  & & & & &\\
LHC(1) & 0.82 & 0.5 & 0.124 & 0.02 & 17.34 & 14.66 \\
  &  & & & & &\\
\hline
\end{tabular}
\end{table}

\subsection{\ud{Temporal Evolution}} The thermally equilibrated
QGP produced at the instant $t_i = \tau_0$ undergoes rapid expansion
(accompanied with cooling) while partonic reactions tend to drive
the plasma towards chemical equilibrium. In Bjorken's boost-invariant
longitudinal expansion scenario the fugacities and temperature
are known~\cite{biro} to evolve through the following master
rate equations :
\be
&&\frac{\dot{\lambda_g}}{\lambda_g} + 3 \frac{\dot{T}}{T} +\frac{1}{\tau}
=R_3 \left(1-\lambda_g\right) - 2 R_2 \left(1-\frac{\lambda_g^2}{\lambda_q^2}
\right),\nonumber\\
&&\frac{\dot{\lambda_q}}{\lambda_q} + 3 \frac{\dot{T}}{T} +\frac{1}{\tau}
=R_2 \frac{a_1}{b_1}\left(\frac{\lambda_g}{\lambda_q} -
\frac{\lambda_q}{\lambda_g}\right),\nonumber\\
&&{\left(\lambda_g+\frac{b_2}{a_2} \lambda_q\right)}^{3/4} T^3 \tau = {\rm{\mbox
{const}}}
\ee
Here $\tau$ is the medium proper time, $\lambda_q$ the quark fugacity,
$N_f$ the number of flavours, and remaining symbols are defined by
\be
&&R_2=0.5 n_g \langle v \sigma_{gg \longrightarrow q \bar{q}} \rangle, \quad
R_3=0.5 n_g \langle v \sigma_{gg \longrightarrow ggg} \rangle \nonumber\\
&&a_1=16 \zeta(3)/\pi^2,\quad  a_2=8\pi^2/15 \nonumber\\
&&b_1=9\zeta(3)N_f/\pi^2,\quad b_2=7\pi^2N_f/20
\ee
Their solutions  on the computer yield the functions $T(t)$,
$\lambda_g(t)$, $n_g(t)$ in terms of the fireball time $t$. The 
lifetime (or freeze-out time) $t_{\rm{life}}$ of the plasma
is the instant when the temperature drops to $T(t_{{}_{\rm{life}}}) = 200$
MeV, say.

\section{Flux-weighted Rate}
\subsection{\ud{Preliminaries}} Next, the question of applying
statistical mechanics to gluonic break-up of the $J/\psi$ becomes
relevant. In the fireball frame consider a $\psi$ meson of mass
$m_\psi$, four momentum $p_\psi$, three velocity $\vec{v}_\psi
=\vec{p}_\psi/p_\psi^0$, and dilation factor $\gamma_\psi = p_\psi^0/
m_\psi$. If $q$ is the gluon four momentum measured in $\psi$ meson
rest frame then by Lorentz transformations
\be
&&K^0 =\gamma_\psi (q^0 +\vec{v}_\psi\cdot \vec{q})
=\gamma_\psi q^0 (1 + \mid \vec{v}_\psi \mid \cos \theta_{q \psi})
\nonumber\\
&& d^3 K = ( K^0/q^0)~d^3 q
\ee
where $\theta_{q\psi}$ is the angle between $\hat{q}$ and $\hat{v}_\psi$
unit vectors.

The invariant quantum mechanical dissociation rate for $g-\psi$
collision can be written compactly as
\be
\Gamma = v_{{}_{\rm{rel}}}~\sigma
\ee
where $v_{\rm{rel}}$ is the relative flux and $\sigma$ the
cross section written in any chosen frame. Its thermal average
over gluon momentum in fireball frame reads
\be
\langle \Gamma \rangle = \frac{16}{n_g} \int \frac{d^3q}{{(2\pi)}^3}
\frac{K^0}{q^0}~\Gamma~f
\ee
with $f$ being the distribution function already encountered
in Eq.(1).

\subsection{\ud{Xu Procedure}} Xu et al.~\cite[Eq.(8)]{xu} worked
with their relative flux $v_{\rm{rel}}^{\rm{Xu}} = q^0/(K^0 \gamma_\psi)$ 
in the fireball frame, but unfortunately the
cross section $\sigma_{{}_{\rm{Rest}}}$ was in $\psi$ meson rest
frame based on the standard QCD value~\cite{BP}
\be
&&\sigma_{{}_{\rm{Rest}}} = B {(Q^0-1)}^{3/2}/ {Q^0}^5 ~;~ q^0 > \epsilon_\psi \\
&&Q^0 = \frac{q^0}{\epsilon_\psi}~,~B = \frac{2 \pi}{3} 
{\left(\frac{32}{3}\right)}^2
\frac{1}{m_c {(\epsilon_\psi m_c)}^{1/2}}
\ee
where $\epsilon_\psi$ is the $J/\psi$ binding energy and $m_c$ the charm
quark mass. Insertion into Eq. (11) led them to
\be
\langle \Gamma^{\rm{Xu}} \rangle = \frac{16}{n_g^{\rm{Xu}}}
\int\frac{d^3 q}{{(2\pi)}^3} \frac{1}{\gamma_\psi} \sigma_{{}_{\rm{Rest}}}~
\lambda_g \sum_{n=1}^\infty e^{-nK^0/T}
\ee
where the approximate $f^{\rm{Xu}}$ given by Eq.(2) has been recalled.
The simple angular integration over $d \cos\theta_{q\psi}$ can be done
by taking the polar axis along $\hat{v}_\psi$ to yield
\be
\langle \Gamma^{\rm{Xu}} \rangle &=& 
\frac{8 \epsilon_\psi^3 \lambda_g}{\pi^2 \gamma_\psi n_g^{\rm{Xu}}}
\sum_{n=1}^\infty \int_1^\infty dQ^0 
{Q^0}^2 \sigma_{{}_{\rm{Rest}}} e^{-C_n Q^0} 
\left( \frac{\sinh D_nQ^0}{D_nQ^0} \right) \\
&=&\frac{4 \epsilon_\psi^2 \lambda_g T}{\pi^2 \gamma_\psi^2 \mid \vec{v}_\psi\mid
n_g^{\rm{Xu}}}
\sum_{n=1}^\infty \frac{1}{n} \int_1^\infty dQ^0~Q^0 \sigma_{{}_{\rm{Rest}}} 
 \left( e^{-{A_n}^- Q^0} - e^{-{A_n}^+ Q^0} \right)
\ee
 
Here  the following abbreviations have been introduced :
\be
\gamma_\psi = p_\psi^0/m_\psi~,~ \vec{v}_\psi = \vec{p}_\psi/p_\psi^0~,
C_n = n \epsilon_\psi \gamma_\psi/T \nonumber\\
D_n = \mid \vec{v}_\psi \mid C_n~,~ A_n^\pm =C_n \pm D_n = C_n 
\left( 1\pm \mid \vec{v}_\psi \mid \right)
\ee
For $J/\psi$  produced in the central rapidity region, Xu et al. have
drawn elaborate curves showing the dependence of
$\langle \Gamma^{\rm{Xu}} \rangle$ on $T$ and $p_T$.

\subsection{\ud{Our Proposal}} We set up our $\Gamma$ entirely in the
$\psi$ meson rest frame where $v_{\rm{rel}}^{\rm{Our}} = c =1$.
Thereby Eq. (11) becomes
\be
\langle \Gamma^{\rm{Our}} \rangle =\frac{16}{n_g^{\rm{Our}}}
\int \frac{d^3q}{{(2\pi)}^3} \gamma_\psi \left( 1 +
\mid \vec{v}_\psi \mid \cos \theta_{q\psi} \right)
\sigma_{{}_{\rm{Rest}}} \sum_{n=1}^\infty
\lambda_g^n e^{-nK^0/T}
\ee
where the exact $f^{\rm{Our}}$ given by Eq.(4) has been recalled.
The slightly complicated angular integration over $d \cos\theta_{q\psi}$
can be performed by choosing the polar axis along $\hat{v}_\psi$
yielding
\be
\langle \Gamma^{\rm{Our}} \rangle  &= &
\frac{8 \epsilon_\psi^3 \gamma_\psi}{\pi^2 n_g^{\rm{Our}}}
\sum_{n=1}^\infty {\lambda_g}^n \int_1^\infty dQ^0~{Q^0}^2 
\sigma_{{}_{\rm{Rest}}} 
e^{-C_n Q^0} \left[ I_0 (\rho_n) - \mid \vec{v}_\psi \mid I_1 (\rho_n)
\right]  \nonumber\\
&=& \frac{4 \epsilon_\psi^2 T}{\pi^2 \mid \vec{v}_\psi\mid n_g^{\rm{Our}}}
\sum_{n=1}^\infty \frac{\lambda_g^n}{n} \int_1^\infty dQ^0~Q^0 
\sigma_{{}_{\rm{Rest}}} \left[ 
\left( 1- \mid \vec{v}_\psi \mid  (1 - \frac{1}{\rho_n}) \right)
e^{-A_n^- Q^0} \right. \nonumber\\
&&\quad \quad \quad \quad \quad \quad \quad \quad~-~\left. 
\left( 1+ \mid \vec{v}_\psi \mid  (1 + \frac{1}{\rho_n}) \right)
e^{-A_n^+ Q^0}  \right]
\ee

where 
\be
&&\rho_n = D_n Q^0 = D_n q^0/\epsilon_\psi~,~I_0 (\rho_n) = 
(\sinh \rho_n)/\rho_n \nonumber\\
&&I_1 (\rho_n) = (\cosh \rho_n)/\rho_n - (\sinh \rho_n)/\rho_n^2
\ee
Clearly the dependence of $\langle \Gamma^{\rm{Our}} \rangle $
on $\lambda_g$ and $\gamma_\psi$ is more involved than that
of $\langle \Gamma^{\rm{Xu}} \rangle $ given by Eqs. (15, 16).

\subsection{\ud{Numerical Work}}
\begin{figure}[h]
\psfig{file=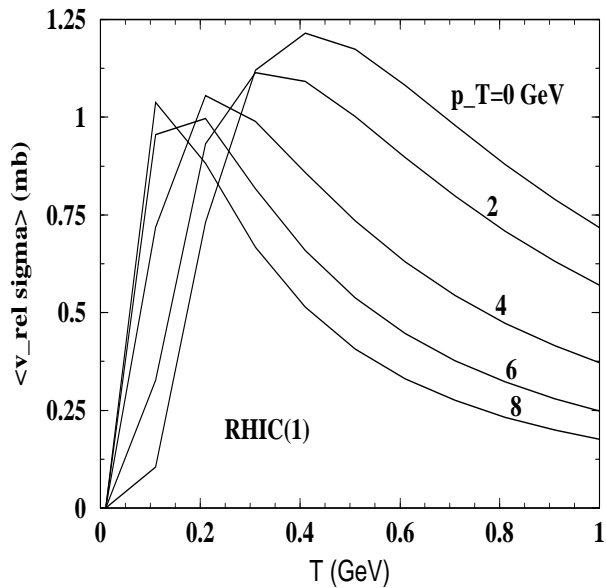,angle=0,height=8cm,width=8cm}
\vskip 0.1in
\caption{The thermal-averaged gluon-$J/\psi$ dissociation
cross section $\langle v_{{}_{\rm{rel}}} \sigma \rangle$ 
as a function of temperature at different transverse momenta
$p_T$ as done by Xu et al.~\cite[Eq.(16)]{xu}.
The initial gluon fugacity is given in Table 1 at
RHIC energy.
}
\end{figure}
The initial thermally-averaged rates $\langle \Gamma^{\rm{Xu}}\rangle $
(Eq.16) and $\langle \Gamma^{\rm{Our}}\rangle $ (Eq.19) are
depicted in Figs. 1, 3 and Figs. 2, 4, respectively. The
physics of dependence of the peak on the temperature and transverse 
momentum has already been discussed in Ref.~\cite{xu}. Here we wish to focus 
attention only on the striking {\em similarity}  between Figs. 1 and 2
inspite of the different fugacities and fluxes employed. For this
purpose, we first go back to the Lorentz transformation (9) and observe
that the Xu  et al relative flux receives dominant contribution from
the antiparallel ($\cos \theta_{q\psi}=-1$) configuration. Indeed,
then
\be
v_{\rm{rel}}^{\rm{Xu}} \equiv \frac{q^0}{\gamma_\psi K^0}
\sim \frac{1}{\gamma_\psi^2 (1-\mid \vec{v}_\psi \mid )}
\sim 1 + \mid \vec{v}_\psi \mid
\ee
Now let us consider the ratio
\be
\frac{\langle \Gamma^{\rm{Xu}} \rangle}{\langle \Gamma^{\rm{Our}} \rangle}
= \left[ \frac{n_g^{\rm{Our}}}{n_g^{\rm{Xu}}} \right] \left[
\frac{\rm{phase~space~integral~of}~v_{\rm{rel}}^{\rm{Xu}}~
\sigma_{{}_{\rm{Rest}}}}{\rm{ phase~space~integral~of~}c~
 \sigma_{{}_{\rm{Rest}}}} \right]
\ee
\begin{figure}[h]
\psfig{file=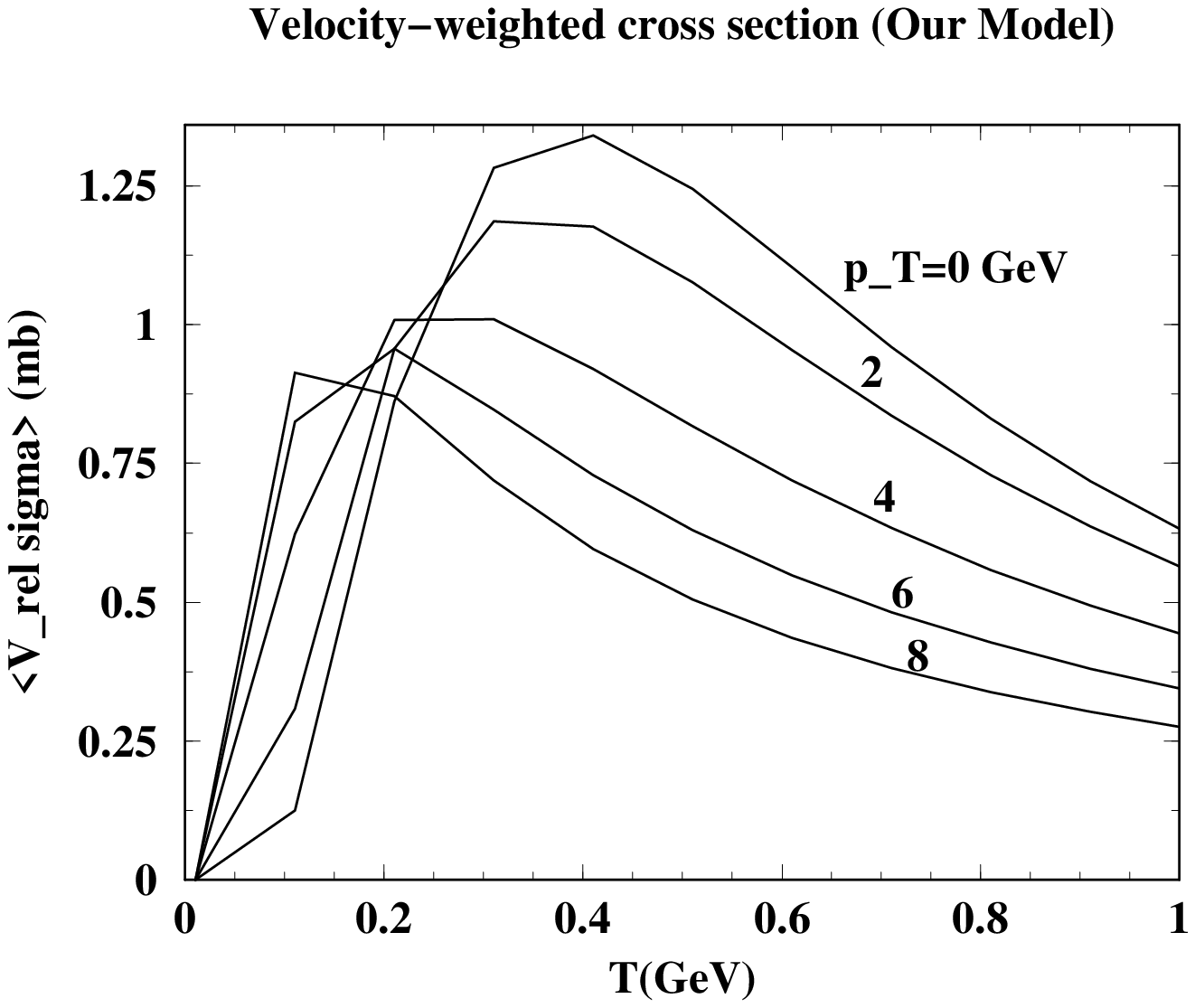,angle=0,height=8cm,width=8cm}
\vskip 0.1in
\caption{Same as in Fig.1, but these curves are obtained
by our Eq.(19).
}
\end{figure}
Due to fugacity effect the number density ratio in Eq.(22) is
somewhat smaller than unity as already mentioned in Sec.2. On the
other hand, since in near antiparallel configuration, the relative
flux $v_{\rm{rel}}^{\rm{Xu}} > c$, hence the ratio of the phase space
integrals is somewhat larger than unity. These two effects tend to
partially compensate each other in Eq.(22) so that the relative
difference between the curves of Figs. 1 and 2 is not more than
about $5-6 \%$. However, the influence of $1+\mid \vec{v}_\psi \mid$
becomes more pronounced at high transverse momentum, causing
noticeable difference between the curves of Figs. 3, 4.
\begin{figure}[h]
\psfig{file=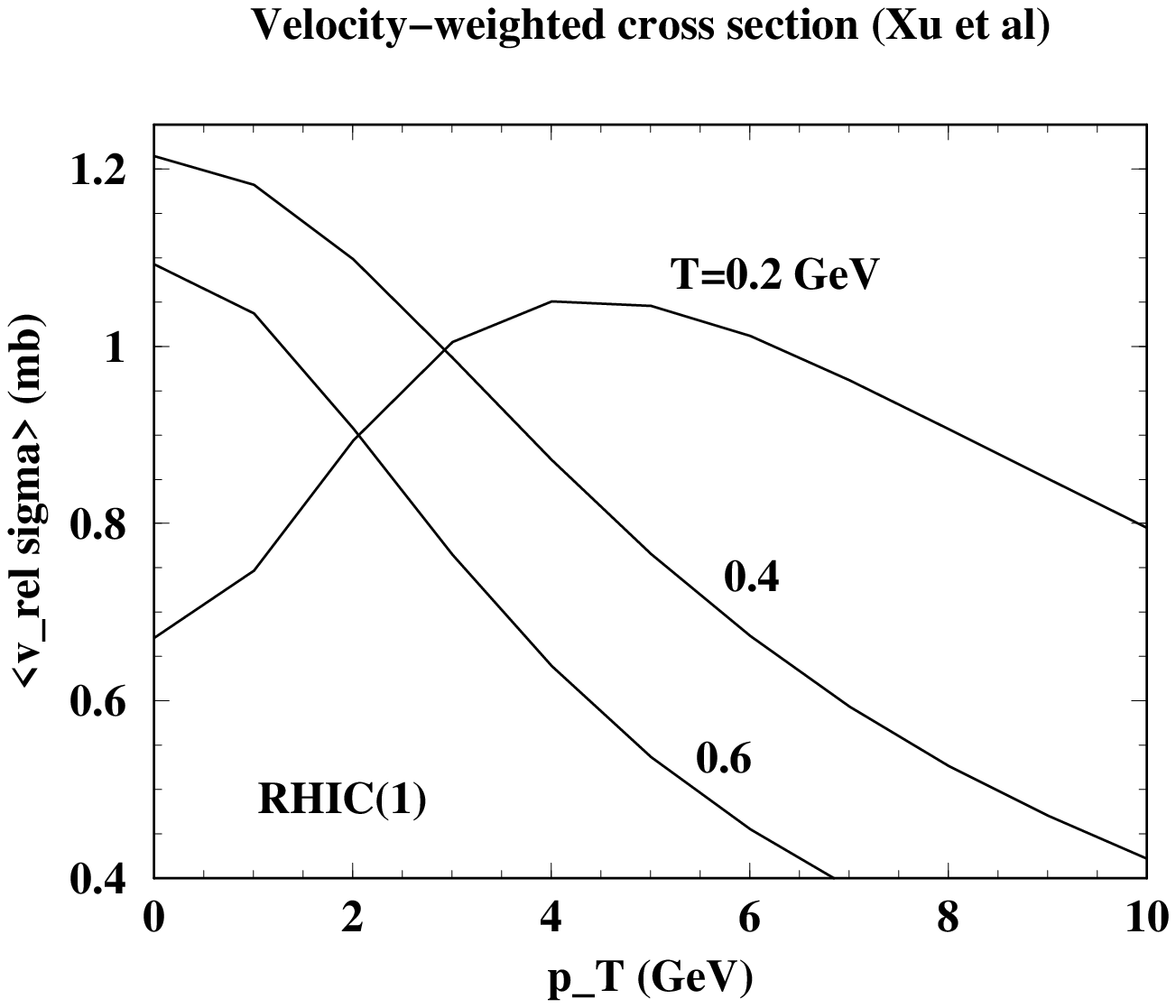,angle=0,height=8cm,width=8cm}
\vskip 0.1in
\caption{The thermal-averaged gluon-$J/\psi$ dissociation
cross section $\langle v_{{}_{\rm{rel}}} \sigma \rangle$ 
as a function of transverse momentum at different 
temperatures as done by Xu et al.~\cite{xu}.
}
\end{figure}
\begin{figure}[h]
\psfig{file=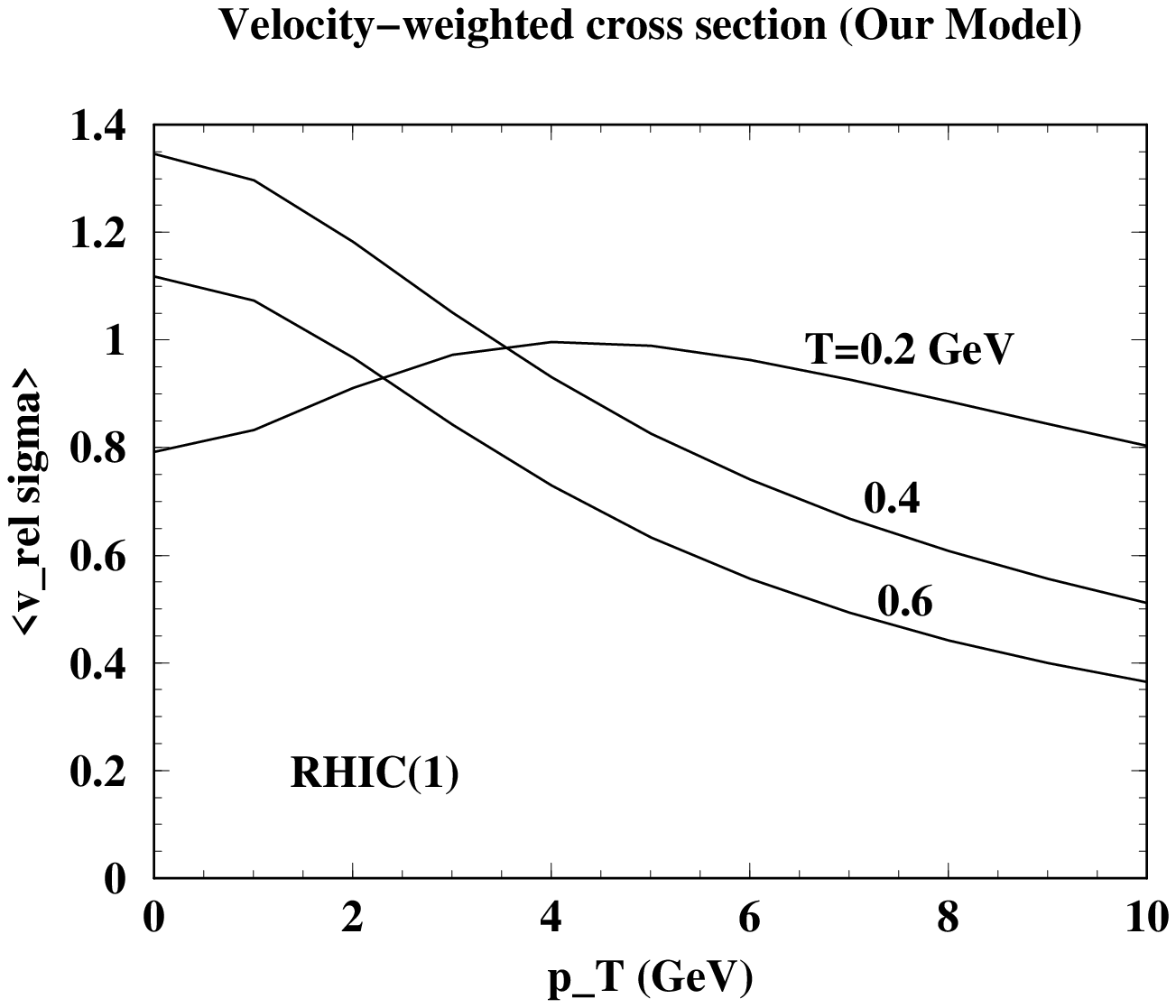,angle=0,height=8cm,width=8cm}
\vskip 0.1in
\caption{same as in Fig.3 except the results are obtained by us.
}
\end{figure}

\section{Survival Probability}
\subsection{\ud{Preliminaries}}
Consider a cylindrical coordinate system in the fireball frame where the
$\psi$ meson was created at the time-space point $(t_I,r_\psi^I, 
\phi_\psi^I)$ with transverse velocity ${\vec{v}_\psi}^T$. The plasma
is supposed to be contained within a cylinder of radius $R$, expanding
longitudinally till the end of its lifetime $t_{{}_{\rm{life}}}$. The $\psi$
meson's trajectory will hit the said cylinder after covering a distance
$d_{{}_{RI}}$ in the time interval $t_{{}_{RI}}$ such that
\be
d_{{}_{RI}} &=& -r_\psi^I \cos \phi_\psi^I + \sqrt {R^2 -{r_\psi^I}^2
\sin^2 \phi_\psi^I} \nonumber\\
t_{{}_{RI}}&=& d_{{}_{RI}}/\mid {\vec{v}_\psi}^T \mid \quad ,
\ee
the full temporal range of interest is obviously
\be
t_I \le t \le t_{II}~;~ t_{II}={\rm{min}} \left( t_i+t_{RI}~,~t_{\rm{life}} 
\right)
\ee
The corresponding survival probability of $J/\psi$ averaged over its
initial production configuration extending over the transverse area $A$
becomes
\be
S(p_T)&=& \int_A d^2 r_\psi^I~ \left( R^2 - {r_\psi^I}^2 \right)
e^{-W} / \frac{}{} \int_A d^2 r_\psi^I \left( R^2 - {r_\psi^I}^2 \right)
\nonumber\\
W&=& \int_{t_I}^{t_{II}} dt~n_g(t)~ \langle \Gamma(t) \rangle
\ee

\subsection{\ud{Xu Procedure}}
Xu et al.~\cite[Eq.(14)]{xu} did not take into account the formation
time of the coulombic bound state, i.e., they chose the instant
of production as
\be
t_I^{Xu} = t_i = \tau_0
\ee
Also, they seem to have used as integration variable the proper
time $\tau_\psi = \left( t - t_i \right) /\gamma_\psi$ measured in $\psi$ 
meson rest frame. This procedure is inconvenient since the gluon number
density $n_g^{Xu}$ was best known in the fireball frame.

\subsection{\ud{Our Proposal}}
We do take into account the formation time $\tau_{{}_F} \sim 0.89 {\rm{fm/c}}$
of the bound state in the $c \bar{c}$ barycentric frame. Remembering 
the dilation factor
$\gamma_\psi$ we choose
\be
t_I^{\rm{Our}} = t_i +\gamma_\psi \tau_{{}_F}
\ee
and retain the fireball time $t$ for integration in Eq.(25).

\subsection{\ud{Numerical Work}}
In Figs.5, 6 the $J/\psi$ survival probability  has been plotted as
a function of the transverse momentum based on the general formula (25).
The solid curve denotes our result using Eq.(27) while the dashed
curve is that of Xu et al employing Eq.(26). Clearly, the $J/\psi$'s
survival chance is much more (i.e.. their suppression is substantially
less) in our case compared to Xu et al's. Its reason can be understood
by examining the integral appearing in Eq.(25) {\em viz.}
\be
W= \int_{t_I}^{t_{II}} dt \left[ {\rm{phase~ space~integral~ of~}}
~v_{\rm{rel}} \sigma ~{\rm{over}}~ f~ {\rm{at~time}}~t \right]
\ee

First, we recall from Sec.2 that $f^{\rm{Our}}< f^{\rm{Xu}}$ due to the fugacity effect.
Next, we know from Eq.(21) that $v_{\rm{rel}} \equiv c < 
v_{\rm{rel}}^{\rm{Xu}}$ due to the flux effect. Finally, Eqs.(26, 27) tell
that the time interval available  for dissociation $t_{II} - 
t_I^{\rm{Our}} < t_{II} -t_I^{\rm{Xu}}$ due to the formation time
effect. These three mechanisms operate {\em cooperatively}
to make $W^{\rm{Our}} < W^{\rm{Xu}}$ resulting in less suppression.

\begin{figure}[h]
\psfig{file=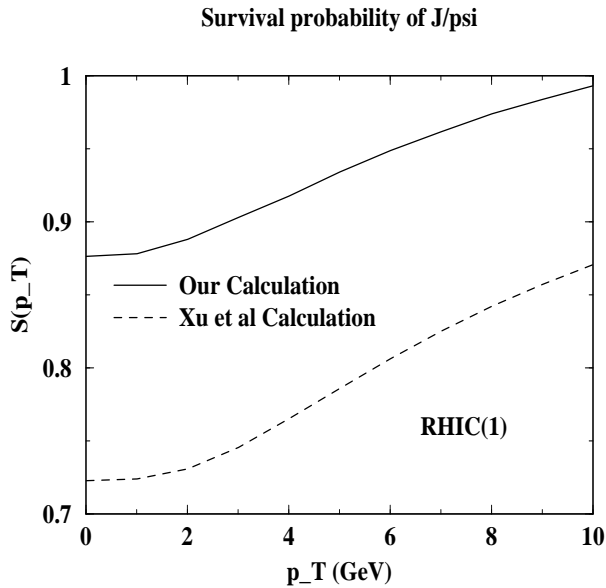,angle=0,height=8cm,width=8cm}
\vskip 0.1in
\caption{The survival probability of $J/\psi$ in an equilibrating 
parton plasma at RHIC energy with initial conditions given in Table 1
~\protect \cite{xu}. The solid curve is our result, while the dashed curve
is the result obtained by Xu et al.~\protect \cite{xu}.
}
\end{figure}
\begin{figure}[h]
\psfig{file=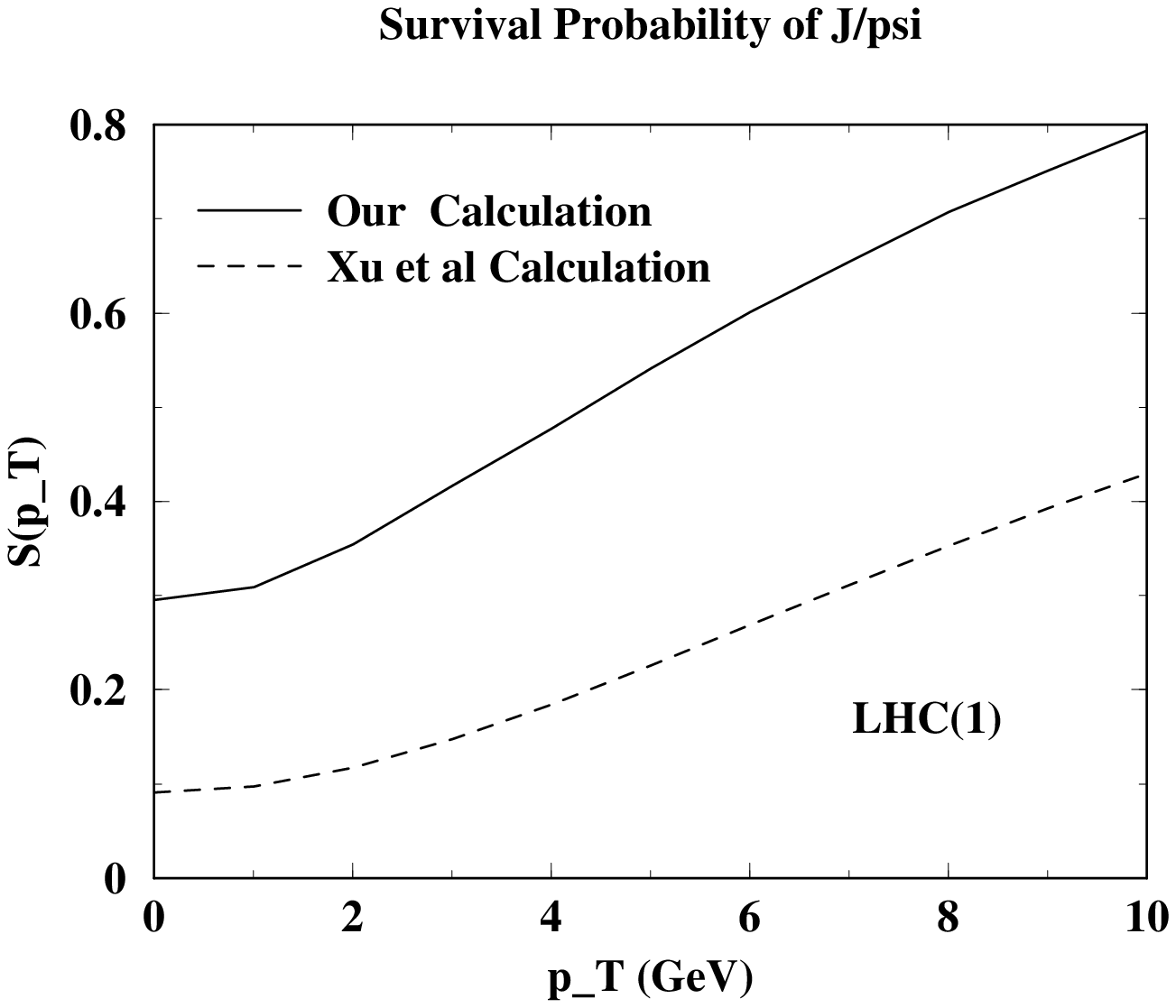,angle=0,height=8cm,width=8cm}
\vskip 0.1in
\caption{Same as in Fig. 5 but it is at LHC energy.
}
\end{figure}

\section{Conclusions}

(i) In this paper we have extended the work of Xu et al~\cite{xu}
concerning the gluonic break-up of the $J/\psi$'s created in an 
equilibrating QGP. Our theoretical formulae on number density
(Eq. 5), flux-weighted cross section (Eq.19), and survival probability
(Eq. 27) are new.

(ii) Our numerical results are also significant as compared to those
of Xu et al. Since gluon {\em fugacity} is less than unity before
chemical equilibration, hence our number density $n_g^{\rm{Our}}(t)$
of hard gluons (which are primarily instrumental in dissociating the
$J/\psi$'s) is lower as shown in Table 1.

(iii) Next, since our $g-\psi$ {\em relative flux} in meson rest frame
is only $v_{\rm{rel}}^{\rm{Our}} =1$ (and not $1+\mid \vec{v}_\psi \mid$
of the fireball frame) hence our thermally-averaged rate 
$\langle \Gamma^{\rm{Our}}(t) \rangle$ is also smaller as depicted
in Fig.4.

(iv) Since we properly take into account the {\em production
time} of the $J/\psi$'s, hence the temporal span available for their
break-up becomes shorter. These three effects act in a cooperative 
manner to reduce substantially the amount of $J/\psi$ suppression
(i.e. to increase noticeably their survival chance $S^{\rm{Our}}
(p_T)$ as demonstrated by Figs. 5, 6.

(v) Apart from possible $ c \bar c$ recombination~\cite{Thews}
another important effect not considered in the present paper is the
{\em transverse} hydrodynamic  expansion of the QGP. Mathematically
such expansion demands that the gluon statistical mechanics
must be done in a local comoving frame, while physically the
temperature will drop more quickly with time. This highly nontrivial
problem is under investigation at present and its results will be
published in a future communication.

\section*{ACKNOWLEDGEMENTS}  
VJM thanks the UGC, Government of India, New Delhi for financial support.
We thank Dr. Dinesh Kumar Srivastava for useful discussions during 
this work.

\newpage
\section*{Appendix}
\renewcommand{\theequation}{A \arabic{equation}}
\setcounter{equation}{0}  
\begin{center}
{\large \bf {$J/\psi$ Suppression Mechanisms Summarized}}
\end{center}
In relativistic heavy ion collision the heavy quark-antiquark pairs
(leading potentially to $J/\psi$ mesons) are produced on a very short
time scale $\simeq 1/2m_c \simeq {10}^{-24}$ sec with $m_c$ being
mass of the charmed quark. The pair develops into the physical resonance
over a formation time $\simeq 0.89 {\rm{fm/c}}$ in its own rest
frame. This $J/\psi$ traverses the deconfined plasma
together with the hadronic matter before leaving the interaction region
to decay into a dimuon which is finally detected. However, this chain of 
events can be prevented via any of the following mechanisms.

Even before the $c\bar c$ bound state is created it may be absorbed by the 
nucleons streaming past (Glauber/normal absorption~\cite{GH}). Or, by the
time the resonance is formed the Debye screening of the colour forces in
the plasma  may be sufficient to kill it~\cite{MS}. Or, an energetic parton
could hit and dissociate the $J/\psi$~\cite{khar}. Or, the Brownian motion of 
the $J/\psi$ through the medium could cause its sufficient
swelling/ionization~\cite{langevin}. The extent of suppression will be decided
by a competition between the $J/\psi$ momentum and the rate of
hydrodynamic expansion (with associated cooling) of the plasma~\cite{dipali}.
Of course, the entire above picture will be substantially modified if the
$J/\psi$'s are regenerated via  $c \bar c$ recombination~\cite{Thews}.

\newpage 
\begin{thebibliography}{39}
\bibitem{review} Helmut Satz, Rept.Prog.Phys. {\bf 63}, 1511 (2000).

\bibitem{xu} Xiao-Ming Xu, D. Kharzeev, H. Satz, and Xin-Nian Wang, Phys.
Rev. C {\bf 53}, 3051 (1996).

\bibitem{langevin} B. K. Patra and V. J. Menon, Nucl. Phys. A {\bf 708},
353 (2002).

\bibitem{screen} B. K. Patra, D. K. Srivastava, Phys. Lett. B {\bf 505},
113 (2001).

\bibitem{hijing} X.-N Wang and M. Gyulassy, Phys. Rev. D {\bf 44}, 3501
(1991).

\bibitem{GH} C. Gerschel and J. H\"{u}fner, Phys. Lett. B {\bf 207}, 253 (1988). 
\bibitem{Thews} R. L. Thews, M. Schroedter, and J. Rafelski,
, Phys. Rev. C {\bf 63}, 054905 (2001). 

\bibitem{khar} D. Kharzeev and H. Satz, Phys. Lett. B {\bf 334}, 155 (1994).

\bibitem{MS} T. Matsui and H. Satz, Phys. Lett. B {\bf 178}, 416 (1986).

\bibitem{HD} H. Satz and D. K. Srivastava, Phys. Lett. B {\bf 475}, (2000).

\bibitem{dipali} D. Pal, B. K. Patra, and D. K. Srivastava, Eur. Phys. Jour.
C {\bf 17}, 179 (2000).

\bibitem{dima} D. Kharzeev and H. Satz, Phys. Lett B {\bf 366}, 316 (1996).

\bibitem{KS} J. P. Blaizot and J. Y. Ollitrault, Phys. Lett. B 199
(1987) 499; F. Karsch and H. Satz, Z. Phys. C51 (1991) 209.

\bibitem{klaus} K. Geiger, Phys. Rep. 258 (1995) 237.

\bibitem{sspc} K. J. Eskola, B. M\"{u}ller, and X.-N. Wang, Phys. Lett.
B 374 (1996) 20.

\bibitem{biro} T. S. Biro, E. van Doorn, M. H. Thoma, B. M\"{u}ller, and
X.-N. Wang, Phys. Rev. C 48 (1993) 1275.

\bibitem{munshi} D. K. Srivastava, M. G. Mustafa, and B. M\"{u}ller, 
Phys. Rev. C 56 (1997) 1064.

\bibitem{eskola} K. J. Eskola, K. Kajantie, P. V. Ruuskanen, and K.
Tuominen, Nucl. Phys. B 570 (2000) 379.

\bibitem{kajantie} K. Kajantie et al., Phys. Rev. Lett. 17 (1997) 3130.

\bibitem{BP} M. E. Peskin, Nucl. Phys. B 156 (1979) 365;
 G. Bhanot and M. E. Peskin, Nucl. Phys. B 156 (1979) 391.

\bibitem{bl} J. P. Blaizot and J. Y. Ollitrault, in {\em Quark Gluon Plasma}
Ed. R. C. Hwa, World Scientific, Singapore, p. 531.

\bibitem{legrand}  L. Gerland,  L. Frankfurt, and M. I.
Strikman, H. St\"ocker, and W. Greiner, Nucl. Phys. A 663 (2000) 1019.

\bibitem{ramona-up} J. Guinon and R. Vogt, Nucl. Phys. B 492 (1997) 301.

\bibitem{khar1} D. Kharzeev and H. Satz, in {\em Quark Gluon Plasma 2}
Ed. R. C. Hwa, World Scientific, Singapore, 1995, p.395.

\end{thebibliography}
\end{document}